\pdfoutput=1

\documentclass[11pt]{article}

\usepackage[]{acl}

\usepackage{times}
\usepackage{latexsym}
\usepackage{amsmath, amssymb}
\usepackage{booktabs}
\usepackage{pifont}

\usepackage{enumitem}
\usepackage{graphicx}
\usepackage{multirow}
\usepackage{diagbox}
\usepackage{tikz}
\usetikzlibrary{shapes.callouts}
\usepackage{xparse}
\usepackage{pgfplots}

\usepackage[T1]{fontenc}

\usepackage[utf8]{inputenc}

\usepackage{microtype}

\usepackage{todonotes}

\title{Searching for a higher power in the human evaluation of MT}

\author{
Johnny Tian-Zheng Wei\thanks{~~~Work done at Microsoft.} \\
University of Southern California \\
\texttt{jtwei@usc.edu}
\And Tom Kocmi \and Christian Federmann
\\ Microsoft \\ 
\texttt{\{tom.kocmi,chrife\}@microsoft.com}
}

\begin{document}
\maketitle


\begin{abstract}
    In MT evaluation, pairwise comparisons are conducted to identify the better system. In conducting the comparison, the experimenter must allocate a budget to collect Direct Assessment (DA) judgments. We provide a cost effective way to spend the budget, but show that typical budget sizes often do not allow for solid comparison. Taking the perspective that the basis of solid comparison is in achieving statistical significance, we study the power (rate of achieving significance) on a large collection of pairwise DA comparisons. Due to the nature of statistical estimation, power is low for differentiating less than 1-2 DA points, and to achieve a notable increase in power requires at least 2-3x more samples. Applying variance reduction alone will not yield these gains, so we must face the reality of undetectable differences and spending increases. In this context, we propose interim testing, an “early stopping” collection procedure that yields more power per judgment collected, which adaptively focuses the budget on pairs that are borderline significant. Interim testing can achieve up to a 27\% efficiency gain when spending 3x the current budget, or 18\% savings at the current evaluation power.
\end{abstract}

\section{Introduction}



In machine translation (MT), pairwise evaluations are conducted to identify the better system over a test domain. MT has long taken intrinsic quality as an object of interest, and assumes it can be determined directly from the output \cite{DBLP:journals/jair/GattK18}. 
Most practitioners accept that human judgments reflect such quality, and take human evaluation as the gold standard \cite{bojar2016ten}. In conducting an evaluation, the experimenter must allocate a budget to collect human judgments, and so evaluation can be an expensive endeavor. No one in the history of MT research has ever been satisfied with the cost or reliability of human evaluation \cite[][inter alia]{graham_baldwin_moffat_zobel_2017, DBLP:conf/acl/LiangCM18, saldias-fuentes-etal-2022-toward}. Likewise, we were keen to find savings, upon the foundation of statistically rigorous inference. 

Evaluation is a noisy process, and we may not expect a repeat experiment to declare the same winners. For one, we may want a holistic answer of the best system over the entire test domain, but we can only evaluate on a small and finite set of input source sentences \cite{koehn-2004-statistical, dror-etal-2018-hitchhikers}. This introduces a sample bias that our conclusion must be wary of. For another, human judgments on the same output may diverge, so we assume that humans are only a noisy reflection of the true intrinsic quality \cite{DBLP:conf/naacl/GrahamBM15}. This introduces additional noise when drawing a conclusion from our observations. Intuitively, using a larger test set or averaging over more human judgments should yield more consistency in pairwise comparison.

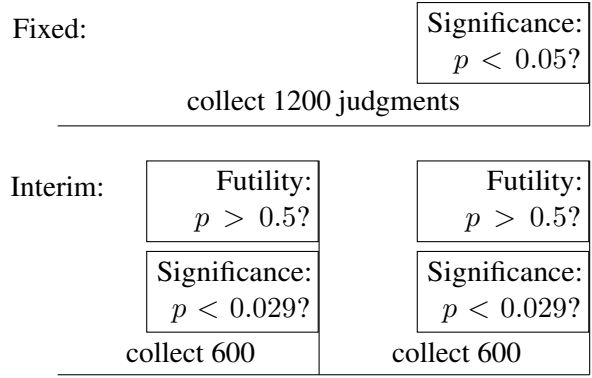
\begin{figure}
    \centering
    \begin{tikzpicture}
    \draw[] (-7,-0.7) -- (0,-0.7) ;
    \node[align=center] at (-3.5,-0.4) {collect 1200 judgments};
    \node[draw, align=right, text width=2cm] at (-1.14, .4){Significance: \\ $p < 0.05$?};
    \draw[] (0,-0.7) -- (0, 0.94) ;
    \node[align=right] at (-7.1,0.6) {Fixed:};

    \node[align=right] at (-7,-1.5) {Interim:};
    \draw[] (-7, -4) -- (0, -4) ;
    \node[align=center] at (-5.25,-3.7) {collect 600};
    \node[align=center] at (-1.75,-3.7) {collect 600};
    
    \node[draw, align=right, text width=2cm] at (-4.7,-1.7) {Futility: \\ $p > 0.5$?};
    \node[draw, align=right, text width=2cm] at (-4.7,-2.9) {Significance: \\ $p < 0.029$?};
    \draw[] (-3.56, -4) -- (-3.56, -1.16);
    
    \node[draw, align=right, text width=2cm] at (-1.14,-1.7) {Futility: \\ $p > 0.5$?};
    \node[draw, align=right, text width=2cm] at (-1.14,-2.9) {Significance: \\ $p < 0.029$?};
    \draw[] (0, -4) -- (0, -1.16);
    
    \end{tikzpicture}
    \caption{A graphical representation of evaluation with different testing procedures. Currently, our evaluation uses fixed testing, and our current budgets (depicted) often result in underpowered comparison (\S\ref{sensitivity_approach}). To get a notable increase in power, we will need to spend more (\S\ref{known_unknowns}), and interim testing is a way to spend efficiently. Interim testing allows for early stopping by trading off power for additional peeks. In MT, such a tradeoff is a favorable and can yield more power per judgment (\S\ref{sec:interim_testing}).}
    \label{fig:fixed_vs_interim}
\end{figure}
Inferential statistics is necessary in MT evaluation to declare ``winning'' MT systems under uncertainty. Basic usage of statistical testing covers the use case of pairwise MT system comparison \cite{mathur-etal-2020-tangled}. After data collection is complete, we can declare significance by computing a p-value (statistical primer in \S\ref{statistical_primer}). When the p-value is low, a real effect is likely to exist. When the p-value is high, repeat experiments will be inconsistent (effectively tossing a coin), and no good decisions can be made even if you used the gold standard Direct Assessment \cite[DA;][]{DBLP:conf/naacl/GrahamBM15} annotation. Significance is the meta-analysis that guards against falsely declaring winners due to noise, with some level of guarantee.

Our work takes the perspective that the basis of solid comparison is in achieving significance. The rate/likelihood an experiment will observe significance is the \textit{power}, and we would like it to be high. At the same time, we would like to minimize human effort and keep costs low. This paper investigates several aspects of the relationship between power and cost in human evaluation:
\begin{enumerate}
    \item \textit{How can we reason about the power of an evaluation?} We recommend a sensitivity perspective to evaluation, where we characterize an evaluation by its minimum detectable effect (MDE), or the smallest pairwise difference the evaluation will reliably yield significance on. By retrospectively analyzing significance in pairwise comparisons, we can derive an empirical MDE. \textbf{Our evaluations can reliably detect up to 2-3 point of DA difference, but comparisons often exhibit even smaller differences.}
    
    \item \textit{How can we notably increase the sensitivity of an evaluation?} With the appropriate power analysis, we can get a rough estimate of the number of samples required to achieve an acceptable sensitivity. To increase the sensitivity to the desired level, we might hope variance reduction techniques can give us the necessary sample efficiency. \textbf{If we wanted half of the past comparisons to reliably achieve significance, we needed at least 2x more samples, far beyond the \textasciitilde{}1.2x sample efficiency variance reduction offers.}
    
    \item \textit{How can we spend more money efficiently?} If we are not satisfied with the power of our current evaluation, increasing the budget and collecting more judgments is necessary. Crucially, if we accept that small differences can't be known, our evaluation can be more efficient by focusing the budget elsewhere. \textbf{We verify that an ``early stopping'' procedure (interim testing) can can achieve up to a 27\% efficiency gain when spending 3x our current budget, or 18\% savings at our current evaluation power.}
\end{enumerate}

\section{Related work}

There is a tradition of using test sets to estimate system performance over the general domain in machine learning \cite{DBLP:books/sp/HastieFT01}. There have been calls for statistically rigorous evaluation in natural language processing using significance testing \cite{dror-etal-2018-hitchhikers}, however its adoption in reporting has been mixed. For a classic task such as part-of-speech tagging, evaluation is generally significant/consistent even for small gains \cite{gorman-bedrick-2019-need}. In MT, even moderate differences in metric gains (e.g. DA, MQM) may not be consistent, so there is a stronger need for significance testing. Historically, MT evaluation has been heavily based on statistical significance \cite{koehn-2004-statistical}.

MDEs have been used to describe the power of experiments in contexts such as education (which program results in increased test scores?) and sociology \cite{doi:10.1177/0193841X9501900504}. \citet{berg-kirkpatrick-etal-2012-empirical} empirically investigate the conventional wisdom that a certain metric gain corresponds to significance (e.g. 0.5 for BLEU). This threshold is exactly an evaluation's MDE. They find that a threshold has strong empirical backing, but a few experimental parameters affect this threshold. In our work, we propose taking a sensitivity perspective to evaluation, and reporting the expected MDE of an experiment instead of the other experimental parameters.

Any statistical technique that reduces the cost of human evaluation is, in another view, improving the power offered by some fixed budget. \citet{DBLP:conf/acl/LiangCM18} first proposed applying control variates to human evaluation. Control variates increase the sensitivity of an evaluation by leveraging information from a metric. This formulation conveniently allows us to analytically understand its performance based on the experimental conditions. In realistic experimental conditions, they found that the sample efficiency gain is at most 20\%, which is in line with results reported in MT \cite{saldias-fuentes-etal-2022-toward}. \citet{mendonca-etal-2021-online} propose using online learning to adaptively spend the evaluation budget on determining the best MT systems. However, their technique lacks in statistical rigor for decision making.

Knowing when to ``early stop'' an evaluation allows us to adaptively spend the budget on difficult pairs and save on easily distinguished pairs. It is known that peeking at the p-value while data collection is ongoing is problematic. Peeking inflates the chance of observing significance and the chance that such significant observation is incorrect \cite{albers2019problem}. While always valid p-values can be calculated that adjust for this error and can be reported at any time, they are mathematically difficult to apply \cite{always-valid-johari}. Interim testing has been used in medical trials, where experimenters have an ethical consideration in stopping the experiment early \cite{10.2307/2530245}. By planning the number of peeks in advance, interim testing can offer rigorous statistical inference while potentially saving time and effort, packaged in an easy to understand technique \cite{lakens_pahlke_wassmer_2021}. Our work investigates whether the tradeoff between power and savings is favorable for MT evaluation.


\section{A primer on inferential statistics} \label{statistical_primer}

We consider pairwise comparisons as the basic unit of evaluation echoing calls from \citet{mathur-etal-2020-tangled} and \citet{kocmi-EtAl:2021:WMT}. Pairwise comparisons are more interpretable than correlations, and more practical for production deployment scenarios. In a pairwise comparison we test the difference between two systems \texttt{A} and \texttt{B}. If you were just to collect a number of DA judgments for each system and declare a winner, a repeat experiment could yield different results due to experimental noise. 

A statistical test guards against making an incorrect conclusion due to experimental noise. To do this, we assume a null hypothesis (that \texttt{A} is better than \texttt{B}) and examine how likely we could have made observed our data under this assumption. There are two outcomes of conducting a test: 
\begin{enumerate}[label=(\roman*)]
    \item there is evidence of a significant difference which rejects the null hypothesis, or
    \item the evidence is insufficient and we are unable to reject the null hypothesis.
\end{enumerate}
In the case of (i), a significance test usually guarantees a false detection rate of at most $\alpha$, where usually $\alpha=0.05$. Therefore, the best outcome of statistical testing is the presence of significance, where our inferences enjoy a low false detection rate. The rate at which we can declare significance is called an experiment's \textit{power} (typically denoted as $1 - \beta$, where $\beta$ is the false negative rate). In pairwise comparison, our evaluation should have an accuracy $(1-\alpha)(1 - \beta)$ against the true, pairwise judgment.\footnote{This pairwise accuracy holds if you assume that different MT systems always have different quality. By randomizing the systems, the null hypothesis will be true exactly half of the time.}

Intuitively, statistical testing can be loosely thought of as reducing the width of two confidence intervals, spaced by the true system difference of \texttt{A} and \texttt{B} \cite{Krzywinski2013}. The power of an experiment is then a function of these three aspects: 
\begin{enumerate}[label=(\alph*)]
\item First, the true system difference plays a role in the power. When the distance between the true scores is large relative to the noise, noise is unlikely to obfuscate the true pairwise ranking of the systems.
\item Second, the variance of human judgment. The larger the variance in a single judgment, the more judgments that will be needed in an average to get a consistent estimate.
\item Finally, the sample size or the budget. The number of judgments you collect shrinks the confidence intervals by a factor of $\sqrt{N}$ from the single judgment variance.
\end{enumerate}
The more judgements you can collect the smaller these confidence intervals will be. When the confidence intervals don’t overlap, the comparison is likely to achieve significance. These three factors all play a role in whether the intervals will be narrow enough.

If we know two of (a), (b), or (c), we can use the appropriate power analysis to deduce the third. Typically, we will observe the (b) human judgment variance, and make a guess at what the true difference (a) would be, to compute what (c) the budget we would have to spend is. When providing estimates for the budget, we would provide estimates under a range of guesses at what the true difference is \cite{card-etal-2020-little}. Alternatively, we may also ask what the minimum detectable effect is for some fixed budget. \citet{wei-jia-2021-statistical} conducted power analysis in MT and found that small differences require an infeasible amount of budget. This gives a hint that most of our MT evaluation is underpowered. Consistently conducting underpowered experiments run the risk of inflating the error rate in significant observations \cite{10.1371/journal.pmed.0020124}.

\section{Dataset}

MT evaluation has an established tradition of conducting human evaluation and releasing public datasets. At the time of writing, the current annotation method of choice in MT is \textit{Direct Assessment} \cite[DA;][]{DBLP:conf/naacl/GrahamBM15, akhbardeh-EtAl:2021:WMT}. Direct Assessment asks annotators to rate a translation's quality on a sliding point scale from 0-100. We study the \textbf{ShipData} presented in \citet{kocmi-EtAl:2021:WMT}, which is the largest human evaluation dataset of pairwise comparisons, accumulated over two years from internal evaluation campaigns at Microsoft Translator. No text is contained i.e. source, references, or outputs, but the raw DA scores are sufficient for our purposes. We focus on this dataset because it is large and often contain comparisons between state-of-the-art systems. It contains 4004 pairwise comparisons between two systems, where each system pair contains about 600 human judgments per system (1200 for both systems).



\begin{figure}
    \centering
    \includegraphics{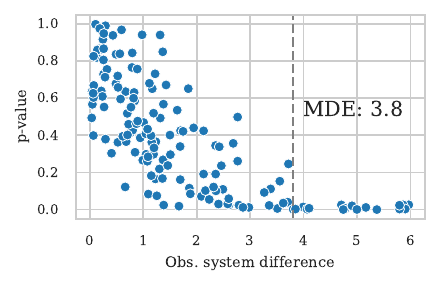}
    \caption{The minimum detectable effect (MDE) is illustrated in the ENU $\rightarrow$ FRA language pair. Each point represents a pairwise comparison conducted for this language pair. When evaluating pairs exhibiting differences larger than the MDE, 95\% of pairs will achieve significance at the $\alpha = .05$ level, which totals to a pairwise accuracy of 90\%. Unfortunately, most pairs are on the left hand side of this line. This is also the case for many other language pairs in the ShipData.}
    \label{fig:obs_mde}
\end{figure}

\section{The sensitivity approach to evaluation} \label{sensitivity_approach}

The basis of solid comparison is significance. Therefore, we need a way to reason about the power of an experiment. In this section, we recommend a sensitivity approach to evaluation, and retrospectively deduce the power of previous evaluations. By looking at the observed effect sizes we can also set a meaningful target power.

\subsection{Minimum detectable effects (MDEs)}

The pairwise evaluation of two MT systems is not a one-size fits all procedure, even though the MT literature uses a consistent annotation method \cite{federmann-2018-appraise}. Rather, an evaluation is our best attempt to answer which MT system is better with the evaluation annotation budget at hand. How much budget to allocate should depend on the circumstantial factors. Statistical inference can give us a probabilistic answer to this question with whatever evidence we are able to collect.


\begin{table}[]
    \centering
\begin{tabular}{r|ccc}
                      & Significant / & Obs. & Median \\
                      & insignificant & MDE                       & difference    \\
\hline
ENU $\rightarrow$ FRA & 30 / 153      & 3.8                       & 1.2             \\
ENU $\rightarrow$ DEU & 19 / 151      & 3.5                       & 0.7             \\
FRA $\rightarrow$ ENU & 3 / 140       & 2.4                       & 0.6             \\
DEU $\rightarrow$ ENU & 27 / 130      & 1.9                       & 0.6             \\
JPN $\rightarrow$ ENU & 78 / 127      & 2.9                       & 3.2             \\
ENU $\rightarrow$ JPN & 40 / 94       & 3.8                       & 1.8             \\
ITA $\rightarrow$ ENU & 2 / 81        & 2.8                       & 0.5             \\
CHS $\rightarrow$ ENU & 30 / 78       & 2.6                       & 1.5             \\
ENU $\rightarrow$ PTB & 28 / 74       & 1.0                       & 0.6             \\
ENU $\rightarrow$ SVE & 31 / 73       & 4.4                       & 1.4            
\end{tabular}
    \caption{Significance and MDE results in the top-10 language pairs (by number of comparisons). Significance is calculated at the $\alpha = 0.05 $ level. Observed MDEs are calculated for 90\% pairwise accuracy. The median system difference is observed from the data. For most language pairs, less than half of the pairs had a significant observation. MDEs are small but most of the system differences appear to be even smaller.}
    \label{tab:obs_mde}
\end{table}

In the best case scenario, a significant result is observed and a winner is declared after the data is collected. However, significance depends on the conditions of the experiment (see \S\ref{statistical_primer}), where the size of the pairwise difference, annotation variance, and number of samples all play a role. The pairwise difference and annotation variance are determined by the annotation method. Since most prefer to use a widely accepted annotation such as Direct Assessment \citep[DA;][]{DBLP:conf/naacl/GrahamBM15}, these are factors we may not be able to change. However, we can increase the budget, and the larger the budget, the more likely we will be able to achieve significance for some fixed difference.

\textbf{We recommend to think about an MT evaluation in terms of its sensitivity. With a fixed budget and annotation method, there is some deducible minimum detectable effect \cite[MDE;][]{doi:10.1177/0193841X9501900504}}, where evaluating differences larger than the MDE will enjoy a comfortable level of power (rate of significance). Alternatively, if we did not observe significance for some experiment, we may suspect that the true difference is likely to be lower than the experiment's MDE. With a sensitivity perspective, our consideration is now to conduct DA evaluations with a budget large enough to exhibit an appropriate MDE. Ideally, our evaluation exhibits an MDE small enough where we believe any smaller differences are not practically meaningful (more in \S\ref{mde_estimation}). Realistically, we would set up an evaluation with MDEs as small as our budgets allow.

\begin{figure}
    \centering
    \includegraphics{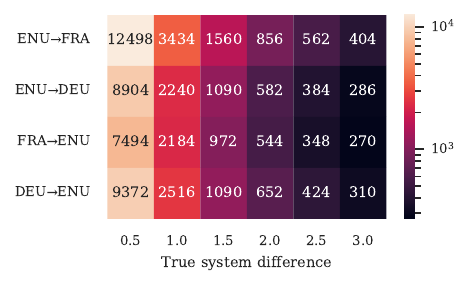}
    \caption{Power analysis for the total number of judgments required to achieve an MDE with 90\% inference accuracy. These figures are calculated through simulation with distributional assumptions on the human scoring function (see \S \ref{mde_estimation}). Compared to the observed MDEs, figures here serve as a lower bound. As the differences decrease linearly, the number of samples required increases exponentially.}
    \label{fig:power_analysis}
\end{figure}

\subsection{Observed MDEs}

In this section, we attempt to retrospectively understand the MDEs/sensitivity of our past evaluations. Refer to Figure \ref{fig:obs_mde} for graphical intuition. We can empirically estimate \cite[as opposed to making assumptions and simulating, see][]{card-etal-2020-little}  an (observed) minimum detectable effect by sorting all the pairs by their observed absolute system difference, and choosing the difference where comparisons with a larger system difference (effect size) will have at least 95\% of experiments showing significance (corresponding to experimental power $1-\beta=0.95$) at a level of $\alpha=0.05$ by the Mann Whitney U (MWU) test. This ensures that at least $(1-\alpha)(1-\beta) \approx 0.9$ of the pairs should be accurate \cite{wei-jia-2021-statistical}. We can interpret this as the threshold at which our experiments will stop being accurate at the 90\% level. 

\begin{table}[]
\centering
\begin{tabular}{ll|ll}
\toprule
 & & Variance & Reducible \\
 & & (std. dev.) & variance \\
 \hline
WMT21 & $\texttt{*-en}$ & 866.2 \small(29.4) & 23.1\% \\
pSQM & $\texttt{zh-en}$ & 683.2 \small(26.1) & 9.8\% \\
pSQM & $\texttt{en-de}$ & 705.4 \small(26.5) & 53.4\% \\
 \bottomrule
\end{tabular}
\caption{Total annotation variance and the reducible proportion of that variance. pSQM scores are provided by \citet{DBLP:journals/corr/abs-2104-14478} and are collected from professional annotators. WMT21 scores are provided by \citet{akhbardeh-EtAl:2021:WMT} and are collected from crowdworkers. pSQM scores are normalized from 0-100 for ease of interpretation. At least half of the variance is irreducible.}
\label{tab:annotator_noise}
\end{table}

\begin{table}[]
    \centering
    \begin{tabular}{c|ccc}
        \diagbox{$\rho$}{} & WMT21 & pSQM\small (zh-en) &  pSQM \small(en-de)  \\
         \hline
        1.0 & 1.30 & 1.20 & 4.33 \\
        0.5 & 1.06 & 1.12 & 3.09 \\
        0.2 & 1.01 & 1.11 & 2.94
    \end{tabular}
    \caption{Data efficiencies for the control variates estimator under different conditions. Each column represents a different condition of reducible variance, instantiated from observed statistics from Table \ref{tab:annotator_noise}. $\rho$ is the correlation of the metric that would be used in the control variates estimator. With the exception in pSQM \texttt{en-de}, variance reduction is far from giving us the 2x-10x multiplier we need.}
    \label{tab:var_reduction}
\end{table}

\textbf{The minimum detectable effects (MDE) are small, but differences between systems are even smaller.} Refer to Table \ref{tab:obs_mde}. Our evaluations have been able to detect up 1 or 2 points of system-level DA difference, but often a third of the comparisons are still not significant. Looking at the density of the differences (see the x-axis in Figure \ref{fig:obs_mde}) we see that most of the pairs exhibit small differences. An immediate consequence is that most of the budget is being spent to declare ties. Most of our comparisons are underpowered, and where the p-value is high the experiments are not much better than a coin toss. The median difference provides a target MDE if we want half of our evaluations to show significance (alternatively, declaring ties in half of the evaluations is acceptable).



\section{Known unknowns} \label{known_unknowns}

Now that we have established a way to reason about experimental power, we conduct power analysis to understand how much more gain we need to improve our power to a desired sensitivity. We investigate whether variance reduction techniques are sufficient, and conclude that the only way forward is to increase the annotation budget.

\subsection{Power analysis for the desired sensitivity} \label{mde_estimation}

As suggested in \citet{card-etal-2020-little}, we can roughly determine the number of samples for a fixed power using simulation. As with any power analysis, we must make some assumptions to estimate the number of samples needed. Here we assume that the judgments for a given system's translation is distributed as $s\sim 100 - \text{Gamma}(k, \theta)$ where $k=\frac{\mu^2}{\sigma^2}$ and $\theta = \frac{\sigma^2}{\mu}$ are fit to match the average mean and variance of a system for that language pair. We choose the use of the Gamma distribution because the resulting scoring distribution is such that most of the scores are high, and the more severe the translation error the more rare it is, which matches what we observe in \citet{kocmi-EtAl:2021:WMT}. We then use the bisection method to determine the integer whose power has the closest match to our desired $\beta$ value. We find that the simulation reasonably matches empirically observed MDEs.

\textbf{Power analysis shows that most pairs needs not a little, but a lot more judgments.} Refer to Figure \ref{fig:power_analysis}. Comparing to the observed MDEs, the power analysis is \textit{optimistic}, where the figures we provide can be seen as a lower bound. Even a reduction of our MDE to 1 point can require up to 2x times more judgments (than originally used in the ShipData). We highlight the fact that as differences get linearly smaller, the number of samples is an exponential growth. The nature of statistical estimation is that smaller differences are increasingly elusive. 

In the search for higher power, we must also keep in mind that arbitrarily small differences require arbitrarily large budgets. Therefore, for modern state-of-the-art comparisons, some differences will be left unknown. We can not fantasize about detecting every single small difference out there just by spending more budget or applying some strong statistical technique (see \S\ref{vr_inadequacies}). Perhaps this may be taken in stride, as mathematicians learned to accept the existence of unprovable theorems nearly a century ago \cite{gdel}. Many other important fields such as domain adaption also grapple with their unknowns \cite{DBLP:journals/jmlr/Ben-DavidLLP10}.

\subsection{Variance reduction is inadequate} \label{vr_inadequacies}


Generally, we assume that a human evaluator scores a segment with the true segment level quality score, plus some noise. If $H(x)$ is the human scoring function on system translations $x$, there are 2 parts to the scoring variance. We can decompose the variance of $H$ to \begin{align} \label{law_of_tv}
    \text{Var}(H(x)) &= \mathbb{E}[\text{Var}(H(x) | x)] \\
    &+ \text{Var}(\mathbb{E}[H(x) | x]) \nonumber
\end{align}
by the law of total variance. The first part is the variance of the true translation quality scores, capturing the real difference in quality across output translations, and the second part is the rest of the variance. The second term, which we broadly term annotator noise, can include annotator biases, preferences, and even mood.

Using repeat judgments we can estimate the second term (annotator noise), which is similar to an inter-annotator agreement \cite{wei-jia-2021-statistical}. Since the ShipData doesn't contain any repeat judgments, we provide estimate of the second term from a few similar datasets \cite{akhbardeh-EtAl:2021:WMT, DBLP:journals/corr/abs-2104-14478}. Refer to Table \ref{tab:annotator_noise}. In designing variance reduction techniques, we usually leverage metric scores to reduce the first term, but not annotator information to reduce the annotator noise (second term), as it is too difficult \cite{saldias-fuentes-etal-2022-toward}.

With variance reduction (VR) techniques, we can achieve a higher power with the current budget by leveraging side information \cite{mcbook}. However, VR is not arbitrarily powerful, and its effectiveness is constrained by the amount of reducible variance present, and how much of the reducible variance you can actually reduce. Here, we look at the control variates technique\footnote{Equal proportion stratified sampling is a special case of control variates, so these results also apply \cite{mcbook}. Any technique which uses a metric to bin outputs, where the same number of outputs are sampled for scoring within each bin, are constrained by these results as well.} which leverages the linear information in a metric for the estimation of system quality. The data efficiency in \citet{DBLP:conf/acl/LiangCM18} describes how many times a control variates estimator improves over the regular sample mean estimate, and is characterized by
\begin{equation}
    \text{DE} := \frac{\text{Var}(\hat{\mu}_{\text{mean}})}{\text{Var}(\hat{\mu}_{\text{cv}})} = \frac{ 1 + \gamma}{1 - \rho^2 + \gamma}
\end{equation}
where $\rho$ is the sentence-level Pearson correlation of the metric and
\begin{equation}
    \gamma = \frac{\sigma^2_a }{ \sigma^2_f } = \frac{\mathbb{E}[\text{Var}(H(x) | x)]}{ \text{Var}(\mathbb{E}[H(x) | x])}
\end{equation}
Refer to Table \ref{tab:var_reduction}. \textbf{With the optimistic assumption of a perfect metric, we often only get a \textasciitilde{}1.2x efficiency gain from VR, far from the 2-10x multiplier we need to obtain significant comparisons.} The gains we predict for VR is consistent with the practical results presented in \citet{DBLP:journals/corr/abs-2204-05307}. These reduction techniques work, but is far from achieving what we need, echoing the narrative of \citet{DBLP:conf/acl/LiangCM18}.

\section{Spending effectively} \label{sec:interim_testing}

To have a notable gain in sensitivity, variance reduction alone is inadequate. Therefore, spending is necessary in the search for higher power. This section describes a simple yet statistically rigorous way of ``early stopping'' in a human evaluation campaign. Interim testing adaptively allocates the budget to borderline significant pairs, and can be seen as an efficient way to spend.

\begin{figure}
    \centering
    \includegraphics{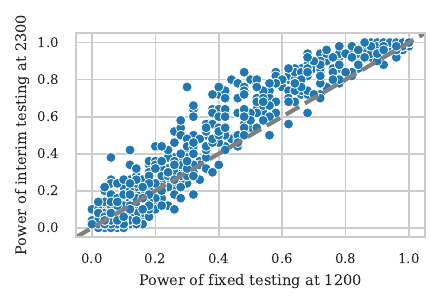}
    \caption{The average power of each pairwise comparison for fixed testing at 1200 against interim-futility testing at 2300. Each point represents a pairwise comparison. When planning for 2300 judgments with interim-futility, the actual amount of judgments collected in our simulation is about 1200. For the same budget, we see that interim-futility testing boosts the power of moderate to high-powered pairs, but drops that of the lower powered pairs.}
    \label{fig:power_fixed_interim}
\end{figure}


\subsection{Peek-a-boo! Planning interim peeks}

Savings can be achieved if we can stop data collection as soon as a result can be concluded. If the experimenter runs the preferred statistical test (at false detection rate $\alpha=0.05$) periodically while data collection is on-going, the final process will have a false detection far higher than the $\alpha$ intended \cite{albers2019problem}. There are a class of sequential sampling techniques, which allow you to test after every single sample while maintaining the false detection rate constant, but are mathematically difficult to apply \cite{always-valid-johari}.

A simpler solution is to use interim sampling and apply a correction for multiple testing \cite{lakens_pahlke_wassmer_2021}. For instance, the Pocock correction \cite{pocock1987analysis} is appropriate when multiple comparisons are made, but we want a false detection to be maintained at a desired $\alpha$.\footnote{Here's why we need a correction: imagine 20 comparisons made at $\alpha=0.05$ where the null hypothesis is true, then the probability of getting at least 1 significant result is actually $1 - 0.95^{20} \approx 0.63$.} Refer to Figure \ref{fig:fixed_vs_interim}. For interim testing, we can plan in advance to collect batches of data, and test between each batch. To maintain a final false detection rate to the fixed procedure, your interim tests must have an $\alpha_0$ appropriately adjusted with the Pocock correction. The downside is that this correction is conservative, and each test has less power.

\begin{figure}
    \centering
    \includegraphics{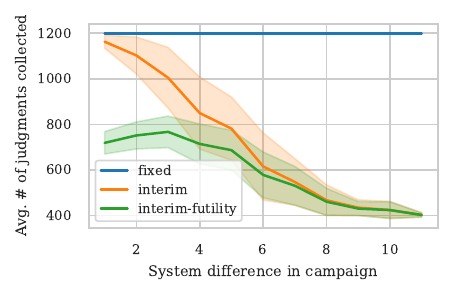}
    \caption{The average number of judgments collected by each sampling method. For interim and interim-futility, 1200 judgments were planned, and the actual judgments collected are strictly less. As the system differences grow larger, both methods have the potential to stop early. For interim-futility, pairs with small differences also incurred less judgments.}
    \label{fig:budget_use}
\end{figure}

At each interim point, we can also stop for futility, or when we see that even in completion of the data collection, we are unable to achieve significance. Practically, there are many ways to set up this stopping rule \cite{lakens_pahlke_wassmer_2021}, but in our simulation we find that a simple heuristic (checking if the $p > 0.5$) works well for our purposes. An alternative view of futility stopping is that we are unwilling to conduct the analysis of the original experiment with the corresponding MDE. 
\begin{figure*}[!ht]
    \centering
    \includegraphics{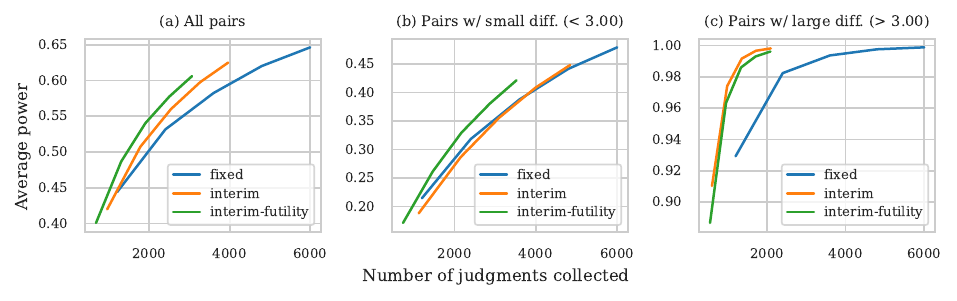}
    \includegraphics[]{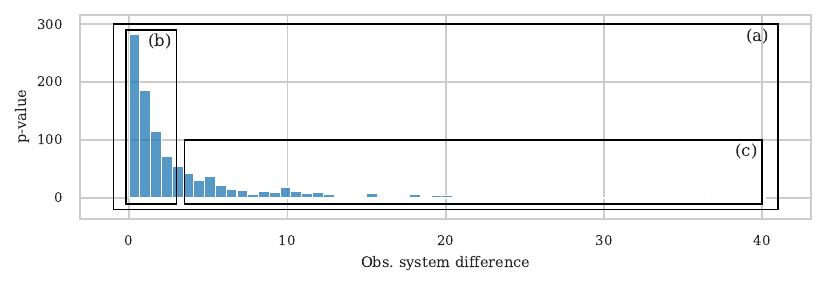}
    \caption{(Top) The average power of each testing procedure across the ShipData for different sized budgets. (a) Shows the average power across all data, and (b) shows it over pairs with large differences and (c) shows it for small differences. (Bottom) The histogram of the true differences in each pairwise comparison. These are true differences due to the simulation we used to test these procedures. Interim-futility is most favorable by average power in (a), (b) and (c). Interim testing is weaker in (c) due to its stricter significance threshold. }
    \label{fig:power_performance}
\end{figure*}

\subsection{Experimental setup}

We compare three different kinds of testing methods. Refer to Figure \ref{fig:fixed_vs_interim}.
\begin{itemize}
    \item \textbf{Fixed} testing is most commonly used in evaluation. In fixed testing, the annotation budget is spent all at once, and the statistical test is performed at the end. The advantage of fixed testing is that only statistical test is performed with the highest (least conservative) alpha threshold (e.g. is $p < 0.05$?).
    
    \item \textbf{Interim} testing plans to spend the budget in equal sized steps, with an interim analysis between each step. If significance is observed at any point, the data collection is terminated. We always plan for 3 peeks, and use the Pocock correction (e.g. is $p < 0.0221$? at each peek). While the testing threshold is lower (more conservative), the savings obtained from some pairs can be used on others, by planning more judgments for all pairs.
    
    \item \textbf{Interim-futility} is the same as interim testing but also applies a futility stopping rule at each analysis. If $p > 0.5$ then the experiment is terminated early. Futility stopping does not affect the false detection rate so it does not need to be adjusted. Futility stopping results in strictly less power, but the savings can be used elsewhere, by planning more judgments.
\end{itemize}

To benchmark these testing procedures against each other, we simulate data collection from the pairs in the ShipData by sampling with replacement. For each pair we simulate each testing procedure 1000 times and record the number of times the procedure is able to achieve significance. For all tests we use the Mann Whitney U test \cite[standard to machine translation;][]{akhbardeh-EtAl:2021:WMT} with a testing threshold of $\alpha = 0.05$. Within the ShipData, each pair only has about 1200 judgments, from which we often oversample. We note that this is our best faith attempt to study these testing methods in the large budget regime, and actual benchmarking would require infeasible cost, so the simulation can serve as our best synthetic testbed.

\subsection{Results}


Refer to Figure \ref{fig:budget_use}. For a fixed sampling procedure, the number of samples collected is constant for every effect size. This can be inefficient as pairs with large differences do not need as many judgments to declare significance. Interim testing is adaptive; as the differences get larger, interim testing can declare significance at an early step. For interim-futility, less judgments are also collected for the pairs with the smallest differences, where early steps may declare futility. We will later see that the interim-futility behavior is most favorable. 

Refer to Figure \ref{fig:power_fixed_interim}. When comparing fixed and interim-futility, we compare two procedures that spend the same budget. Since interim sampling spends more on borderline pairs, the power for pairs with moderate to high differences increases. Savings are made on pairs with both large and small differences, with small difference pairs having a decrease in power. We highlight that interim-futility is a different kind of testing. \textbf{While the use of fixed testing seeks to best detect every difference no matter how small, the use of interim-futility prioritizes the pairs that have borderline significant differences.}

Refer to Figure \ref{fig:power_performance}(a). The main metric we benchmark these methods is by the average power, or the number of significant comparisons over all the ShipData. When comparing over all pairs, interim testing has slightly better performance, but interim-futility gives considerable gains even at current budget sizes. Our results show that to attain the fixed testing power at 1200, interim testing only needed to spend 990 judgments per comparison, which is an 18\% saving\footnote{All the results in this paragraph are derived using linear interpolation, akin to using a ruler on Figure \ref{fig:power_performance}.}. The advantage of interim sampling over fixed sampling is even more pronounced when we are spending large budget sizes, where we can gain 28\% savings at 3600 judgments (3x). Refer to Figure \ref{fig:power_performance}(b). When testing small differences interim sampling is worse than fixed sampling, as it has a stricter significance threshold. However, interim-futility is able to stop on pairs with little hope and prioritize the borderline significant pairs. Refer to Figure \ref{fig:power_performance}(c). On pairs with large differences interim sampling is best, with interim-futility achieving similar performance. For large differences futility stopping should rarely trigger, so the two methods should be similar.

\textbf{We want to highlight that the distribution of the differences is key to the success of the interim-futility testing procedure.} Since most of the pairs are concentrated either in the dense region of small differences or in the long tail of large differences, these are areas where interim-futility can early stop. Compared to fixed testing, interim-futility will be able to make savings here to spend elsewhere. Crucially, the application of futility stopping also requires a change in our evaluation mindset, as we must be willing to accept that some small differences are not worth detecting. If we can make this change, then interim-futility is most favorable in terms of average power.

\section{Limitations}


The most important assumption of our work is in the use of Direct Assessment (DA). While our methods can generalize to any real valued judgment, we analyzed DA because of its widely recognized, gold standard status in MT evaluation. DA is a particularly noisy judgment, and so the power and variance reduction results are pessimistic. However, we believe that the study of annotation will be the most important direction in MT evaluation. 

Let's take \citet{DBLP:journals/corr/hassan-2018}, where one of the first claims of MT-human parity was made. By their evaluation, which was conducted according to the community standard, no significant difference was found between human and machine translations with a reasonable budget, and so a tie was declared. \citet{toral-etal-2018-attaining} reassesses this claim, and essentially presents a series of alternative evaluations and observe significant differences that contradict with \citet{DBLP:journals/corr/hassan-2018}. This is just one of many studies which compels an alternative evaluation with qualitative insight, overriding quantitative evidence \cite{DBLP:conf/emnlp/LaubliS018, DBLP:journals/jair/LaubliCNSST20, DBLP:journals/corr/abs-2104-14478}.

Our perspective is that significance is only one pillar of MT evaluation. It is our hope that the analyses in this work will further our understanding of significance and evaluation power. However, the second pillar of MT evaluation is in the annotation method. While power is quantitative, the study of annotation methods will be qualitative. Going forward, understanding how we can change the annotation method to increase the power will be crucial. We will need good qualitative understanding to be able to move away from DA and establish new gold standards.

In addition, we showed that interim testing is only effective for pairwise comparison. Future work should look to make savings in the leaderboard styled evaluation of WMT. This may come in the form of generalizing interim sampling for multiple comparisons or formalizing the bandit results from \citet{mendonca-etal-2021-online} in terms of statistical inference.



\section{Conclusion}

Our work is motivated by the cost of human evaluation in machine translation. Before searching for a higher power from our current budget, we determined how much more power was necessary. In doing so, we recommend taking a sensitivity approach to evaluation. From here we came to the conclusion that to achieve the power/sensitivity necessary, variance reduction alone would be insufficient, and spending is our only option. If we decide to allocate larger budgets, interim testing is a more effective way to spend, which can yield 18\% savings at the current evaluation power, or 27\% savings at 3x the original budget.

\section{Acknowledgments}

The first author completed this work during the remote internship program at Microsoft Research. Feedback from Arul Menezes, Matt Post, Huda Khayrallah, Roman Grundkiewicz, Hitokazu Matsushita, and other members of the Translator team helped shaped this work. Robin Jia advised the first author on related prior work. Sojhal Bloach graciously provided free housing to the first author in Seattle. Thanks to all who have made this work possible. Finally, I left a part of me in Missoula, Montana. Let's promise to hold onto each other until it loses its meaning.

\bibliography{anthology,custom}
\bibliographystyle{acl_natbib}

\appendix



\end{document}